\documentclass[%
preprint,
pra,
amsmath,amssymb,
aps,
]{revtex4-1}
\usepackage{newtxtext,newtxmath,braket}
\usepackage[dvipdfmx]{graphicx}
\begin{document}
\title{Direct estimation of the energy gap between the ground state and excited state
 with quantum annealing
}
\author{Yuichiro Matsuzaki$^{1}$}\email{matsuzaki.yuichiro@aist.go.jp}
\author{Hideaki Hakoshima$^{1}$}
\author{Kenji Sugisaki$^2$}
\author{Yuya Seki$^{1}$}\author{Shiro Kawabata$^{1}$}
\affiliation{$^{1}$ Device Technology Research Institute, 
National Institute of Advanced Industrial Science and Technology (AIST)
Umezono 1-1-1, Tsukuba, Ibaraki 305-8568 Japan \\
}
\affiliation{$^{2}$ 
Department of Chemistry and Molecular Materials Science, Graduate School of
Science, Osaka City University, 3-3-138 Sugimoto, Sumiyoshi-ku, Osaka 558-8585
 \\
}

\begin{abstract}
Quantum chemistry is one of the important applications 
of quantum information technology. Especially, an
estimation of the energy gap between a ground state and 
excited state of a target Hamiltonian corresponding to a molecule is
 essential. In the previous approach, an energy of the ground state and that of 
the excited state are estimated separately, and the energy gap can be calculated from the subtraction 
between them. Here, we propose a direct estimation of the energy gap between 
the ground state and excited state of the target Hamiltonian with quantum annealing. 
The key idea is to combine a Ramsey type measurement with the quantum annealing. 
This provides an oscillating signal with a frequency of the energy gap, and a Fourier
 transform of the signal let us know the energy gap. Based on typical parameters of 
superconducting qubits, we numerically investigate the performance of our scheme when we estimate 
an energy gap between the ground state and first excited state of the Hamiltonian. 
We show robustness against non-adiabatic transitions between the ground state and first-excited state.
 Our results pave a new way to estimate the energy gap of the Hamiltonian
for quantum chemistry.
\end{abstract}
\maketitle

\section{Introduction}
Quantum annealing (QA) has been studied as a way to solve combinational 
optimization problems~\cite{kadowaki1998quantum,farhi2000quantum,farhi2001quantum}
where the goal is to minimize a cost function. Such a problem is mapped 
into a finding of a ground state of Ising Hamiltonians that contain the information of the problem. 
QA is designed to find an energy eigenstate of the target Hamiltonian by using adiabatic dynamics. 
So, by using the QA, we can find the ground state of the Ising Hamiltonian for the combinational optimization problem.

D-Wave systems, Inc. has
 have realized a quantum device to implement the QA \cite{johnson2011quantum}.
Superconducting flux qubits \cite{orlando1999superconducting,mooij1999josephson,harris2010experimental}
have been used in the device 
for the QA. Since superconducting qubits are artificial atoms, 
there are many degree of freedoms to control parameters
by changing the design and external fields, which is suitable for a programmable device.
QA with the D-Wave machines can be used not only for finding the ground state, but also for 
quantum simulations \cite{harris2018phase,king2018observation}
and machine learning \cite{mott2017solving,amin2018quantum}.

Quantum chemistry is one of the important applications of 
quantum information processing~\cite{levine2009quantum,serrano2005quantum,mcardle2020quantum}, and
it was recently shown that the QA can be also used for quantum chemistry 
calculations~\cite{perdomo2012finding,aspuru2005simulated,lanyon2010towards,du2010nmr,peruzzo2014variational,
mazzola2017quantum,streif2019solving,babbush2014adiabatic}. 
Important properties of molecules can be investigated by the second quantized Hamiltonian of the molecules. 
Especially, the energy gap between the ground state and excited states is essential information for 
calculating optical spectra and reaction rates
~\cite{serrano2005quantum}. 
The second quantized Hamiltonian can be mapped into the Hamiltonian of 
qubits~\cite{jordanwigner,bravyi2002fermionic,aspuru2005simulated,seeley2012bravyi,tranter2015b}. 
Importantly, not only the ground state 
but also the excited state of the Hamiltonian can be prepared by the QA \cite{chen2019demonstration,seki2020excited}. By measuring suitable observable on 
such states prepared by the QA, we can estimate the eigenenergy of the Hamiltonian. In the conventional approaches, 
we need to perform two separate experiments to estimate an energy gap between the ground state and the excited state. 
In the first (second) experiment, we measure the eigenenergy of the ground (excited) state prepared by the QA. From the subtraction between the estimation of the eigenenergy of the ground state and that of the excited state, we can obtain the information of the energy gap \cite{seki2020excited}.

Here, we propose a way to estimate an energy gap between the ground state and excited state in a more direct manner. 
The key idea is to use the Ramsey type measurement where a superposition between the ground state and excited state 
acquires a relative phase that depends on the energy gap \cite{ramsey1950molecular}. By performing the Fourier transform of the signal from the 
Ramsey type experiments, we can estimate the energy gap. We numerically study the performance of our protocol to estimate 
the energy gap between the ground state and first excited state. We show robustness of our scheme against non-adiabatic 
transitions between the ground state and first excited state.

\section{Estimation of the energy gap between the ground state and excited state based on the Ramsey type measurement
}

\begin{figure}[bhtp]
  \centering
  \includegraphics[width=16cm]{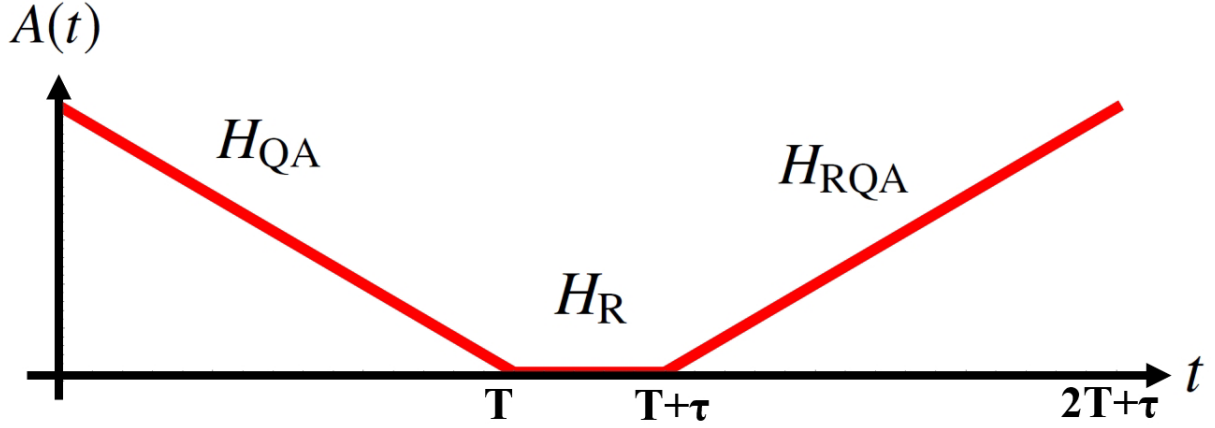}
  \caption{
An external control parameter $A(t)$ of our time-dependent Hamiltonian 
$H(t)=A(t)H_{\rm{D}}+(1-A(t))H_{\rm{P}}$ where $H_{\rm{D}}$ denotes
the driving Hamiltonian 
and $H_{\rm{P}}$ denotes the target (problem) Hamiltonian. 
With a time period of $0\leq t \leq T$, we have the Hamiltonian $H_{\rm{QA}}$ that is used in the standard QA.
With the next time period of $T \leq t \leq T+\tau $, we have the Ramsey time Hamiltonian $H_{\rm{R}}$
where the quantum state acquires a relative phase induced from the energy gap.
In the final time period of $T+\tau \leq t \leq 2T+\tau $, we have the Hamiltonian 
$H_{\rm{RQA}}$ which is used in a reverse QA. By using the dynamics induced by these Hamiltonians, we can estimate the energy gap of the target Hamiltonian.
}\label{aatfigure}
\end{figure}

We use the following time-dependent Hamiltonian in our scheme
\begin{eqnarray}
 H&=&A(t)H_{\rm{D}}+(1-A(t))H_{\rm{P}}\nonumber \\
A(t)&=&\left\{ \begin{array}{ll}
1 -\frac{t}{T}& (0\leq t \leq T) \\
0 & (T \leq t \leq T +\tau ) \\
\frac{t-(T+\tau )}{T} & (T+\tau \leq t \leq 2T+\tau )
\\
\end{array} \right.
\end{eqnarray}
where $A(t)$ denotes an external control parameter (as shown in the Fig. \ref{aatfigure}), $H_{\rm{D}}$ denotes the driving Hamiltonian that is typically chosen as the transverse magnetic field term,
and $H_{\rm{P}}$ denotes the target (or problem) Hamiltonian whose energy gap we want to know.
This means that, depending on the time period, 
we have three types of the Hamiltonian as follows
 \begin{eqnarray}
  H_{\rm{QA}}&=&(1-\frac{t}{T})H_{\rm{D}}+\frac{t}{T}H_{\rm{P}} 
 \nonumber \\
 H_{\rm{R}}&=&H_{\rm{P}}\nonumber \\
 H_{\rm{RQA}}&=&\frac{t-(T+\tau )}{T}H_{\rm{D}}+(1-\frac{t-(T+\tau )}{T})H_{\rm{P}}\nonumber 
 \end{eqnarray}
In the first time period of  $0\leq t \leq T$, 
the Hamiltonian is $H_{\rm{QA}}$, and this
is the same as that is used in the standard QA.
In the next time period of $T \leq t \leq T +\tau$, 
the Hamiltonian becomes $H_{\rm{R}}$, and
the dynamics induced by this Hamiltonian 
corresponds to that of the Ramsey type evolution \cite{ramsey1950molecular} where the superposition
of the state acquires a relative phase depending on the energy gap.
In the last time period of $T+\tau \leq t \leq 2T+\tau$, 
the Hamiltonian becomes $H_{\rm{RQA}}$, and this has a similar form of that
 is used in a reverse QA
\cite{perdomo2011study,ohkuwa2018reverse,yamashiro2019dynamics,arai2020mean}.

We explain the details of our scheme.
Firstly, prepare an initial state of
$|\psi _0\rangle =\frac{1}{\sqrt{2}}(|E_0^{\rm{(D)}}\rangle +|E_1^{\rm{(D)}}\rangle)$
where $|E_0^{\rm{(D)}}\rangle$ ($|E_1^{\rm{(D)}}\rangle$) 
denotes the ground (excited) state of the driving Hamiltonian. 
Secondly, let this state evolve in an adiabatic way
by the Hamiltonian of $H_{\rm{QA}}$
and we obtain a state of
$\frac{1}{\sqrt{2}}(|E_0^{\rm{(P)}}\rangle +e^{-i\theta }|E_1^{\rm{(P)}}\rangle)$
where $|E_0^{\rm{(P)}}\rangle$ ($|E_1^{\rm{(P)}}\rangle$) 
denotes the ground (excited) state of the target Hamiltonian and $\theta $ 
denotes a relative phase acquired during the dynamics. Thirdly, let the state evolve by the Hamiltonian
of $H_{\rm{R}}$
for a time $T\leq t \leq T+\tau $, and we obtain 
$\frac{1}{\sqrt{2}}(|E_0^{\rm{(P)}}\rangle +e^{-i\Delta E \tau 
-i\theta }|E_1^{\rm{(P)}}\rangle)$ where $\Delta E= E_1^{(\rm{P})}-E_0^{(\rm{P})}$ denotes
an energy gap and $E_0^{(\rm{P})}$ ($E_1^{(\rm{P})}$) denotes the eigenenergy of the ground 
(first excited) state of the target Hamiltonian. 
Fourthly, let this state evolve in an adiabatic way
by the Hamiltonian of $H_{\rm{RQA}}$ from $t=T+\tau $ to $T$,
and we obtain a state of
$\frac{1}{\sqrt{2}}(|E_0^{\rm{(D)}}\rangle +e^{-i\Delta E \tau 
-i\theta '}|E_1^{\rm{(D)}}\rangle)$ 
where $\theta '$ denotes 
a relative phase acquired during the dynamics. Fifthly, we readout the state by using a projection operator of
$|\psi _0\rangle \langle \psi _0|$, and the projection probability is 
$P_{\tau }=\frac{1}{2}+\frac{1}{2} \cos (\Delta E \tau +\theta ')$, which is 
an oscillating signal with a frequency of the energy gap.
 Finally, we repeat the above five steps by sweeping $\tau $, and obtain several values of $P_{\tau }$.
We can perform the Fourier transform of $P_{\tau }$ such as 
\begin{eqnarray}
 f(\omega )= \sum_{n=1}^{N}(P_{\tau }-\frac{1}{2})e^{-i\omega \tau _n}
\end{eqnarray}
where $\tau _n= t_{\rm{min}}+\frac{n-1}{(N-1) }(t_{\rm{max}} - t_{\rm{min}})$ 
denotes a time step,  $t_{\rm{min}}$ ($t_{\rm{max}}$)
denotes a minimum (maximum) time to be considered, 
and $N$ denotes the number of the steps. The peak in $f(\omega )$ shows the energy gap $\Delta E$.

\begin{figure}[bhtp]
  \centering
  \includegraphics[width=16cm]{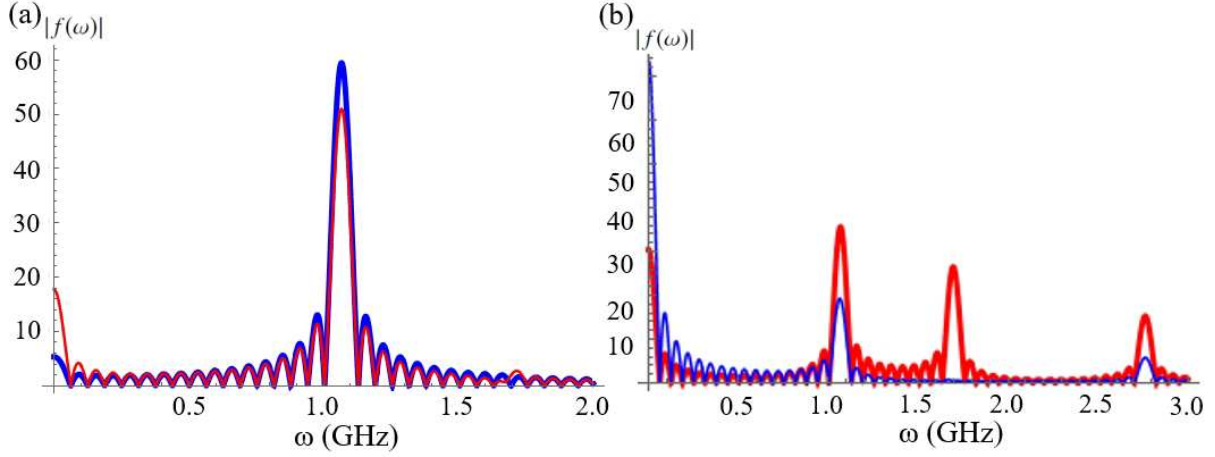}
  \caption{Fourier function against a frequency. Here, we set parameters as $\lambda _1/2\pi =1$ GHz, $g/2\pi =0.5$ GHz, $\omega _1/2\pi = 0.2$ GHz, $\omega _2/\omega _1=1.2$, $g'/g=2.1$, $\lambda _2/\lambda _1 =10.7$, $L=2$,
$N=10000$, $t_{\rm{min}}=0$, $t_{\rm{max}}=100$ ns. 
(a) We set $T=150 \ (75)$ ns for the blue (red) plot, 
where we have a peak around $1.067$ GHz, which corresponds to 
the energy difference between the ground state and first excited state of the target Hamiltonian.  We have another 
peak around $\omega =0$, and this comes from non-adiabatic transition during the QA.
(b) We set  $T=37.5$($12.5$) ns for the blue (red) plot. We have an additional peak around $1.698$ GHz ($2.7646$ GHz),
which corresponds to 
the energy difference between the first excited state (ground state) and second excited state  of the target Hamiltonian.  
}\label{dcodmr}
\end{figure}

To check the efficiency, we perform the numerical simulations to estimate the energy gap between the ground state and first excited state, 
based on typical parameters for superconducting qubits. We choose the following Hamiltonians
\begin{eqnarray}
 H_{\rm{D}}&=&\sum_{j=1}^{L}\frac{\lambda _j}{2}\hat{\sigma }_x^{(j)}\nonumber \\
H_{\rm{P}} &=&\sum_{j=1}^{L} \frac{\omega _j}{2}\hat{\sigma }_z^{(j)}
+\sum_{j=1}^{L-1}g \hat{\sigma }_z^{(j)}\hat{\sigma }_z^{(j+1)}
+g'(\hat{\sigma }_+^{(j)} \hat{\sigma }_-^{(j+1)} + \hat{\sigma }_-^{(j)} \hat{\sigma }_+^{(j+1)} )
\end{eqnarray}
where $\lambda _j$ denotes the amplitude of the transverse magnetic fields of the $j$-th qubit,
$\omega _j$ denotes the frequency of the $j$-th qubit, and $g$($g'$) denotes the Ising (flip-flop) 
type coupling strength between qubits. 

We consider the case of two qubits, and 
the initial state is $|1\rangle |-\rangle $ where $|1\rangle $ ($|-\rangle $)
is an eigenstate of $\hat{\sigma }_z$($\hat{\sigma }_x$)
with an eigenvalue of +1 (-1). In the Fig. \ref{dcodmr} (a), we plot the Fourier function $|f(\omega )|$ against $\omega $ for this 
case.  When we set $T=150$ (ns) or $75$ (ns), 
we have a peak around $\omega = 1.067$ GHz, which corresponds to the energy gap $\Delta E$
of the problem Hamiltonian in our parameter. So this result shows that we can estimate the energy gap by using our scheme.
Also, we have a smaller peak of around $\omega =0$ in the Fig. \ref{dcodmr} (a), 
and this comes from non-adiabatic transitions between the ground state and first excited state. 
If the dynamics is perfectly adiabatic, the population of both the ground state and first excited state should be 
$\frac{1}{2}$ at 
$t=T$.  
However, in the parameters with $T=150$ ($T=75$) ns, 
the population of the ground state and excited state is around 0.6 (0.7) and 0.4 (0.3) at $t=T$, respectively. 
In this case, the probability at the readout step should be modified as 
$P'_{\tau }=a+b \cos (\Delta E \tau +\theta ')$ where the parameters 
$a$ and $b$ deviates from $\frac{1}{2}$ due to the non-adiabatic transitions. This induces the peak of around 
$\omega =0$ in the Fourier function $f(\omega )$. As we decrease $T$, 
the dynamics becomes less adiabatic, and the peak of $\omega =0$ becomes higher 
while the target peak corresponding the energy gap $\Delta E $  becomes smaller as shown in the Fig. 1. 
Importantly, we can still identify the peak corresponding to the energy gap for the following reasons. 
First, there is a large separation between the peaks. 
Second, the non-adiabatic transitions between 
the ground state and first excited state
do not affect the peak
position. So our scheme is robust against the non-adiabatic transition between 
the ground state and first excited state. 
This is stark contrast with a previous scheme that is fragile against such non-adiabatic transitions
\cite{seki2020excited}.

Moreover, we have two more peaks in the Fig. \ref{dcodmr} (b) where we choose 
$T=37.5$($12.5$) ns for the red (blue) plot, which is shorter than that of the Fig. \ref{dcodmr} (a).
The peaks are around $1.698$ GHz and $2.7646$ GHz, respectively.
The former (latter)
peak corresponds to 
the energy difference between the first excited state (ground state) and second excited state.
We can interpret these peaks as follows.
Due to the non-adiabatic dynamics, not only the first excited state but also the second excited state is 
induced in this case. The state after the evolution with $H_{\rm{R}}$ at $t=T+\tau $
 is approximately given as
a superposition between the ground state, the first excited state, 
and the second excited state such as 
$c_0e^{-i E_0^{\rm{(P)}} \tau 
-i\theta _0} |E_0^{\rm{(P)}}\rangle +c_1e^{-i E_1^{\rm{(P)}} \tau 
-i\theta _1} 
|E_1^{\rm{(P)}}\rangle + c_2e^{-i E_2^{\rm{(P)}} \tau 
-i\theta _2}
|E_2^{\rm{(P)}}\rangle$ where $c_{i}$ $(i=0,1,2)$ denote real values and $\theta _i$ ($i=0,1,2$)
denotes 
the relative phase induced by the QA.
So the Fourier transform of the probability distribution obtained from the measurements provides us with three frequencies
such as
$(E_0^{\rm{(P)}}-E_1^{\rm{(P)}})$, $(E_1^{\rm{(P)}}-E_2^{\rm{(P)}})$, and $(E_2^{\rm{(P)}}-E_0^{\rm{(P)}})$.
In the actual experiment, we do not know which peak corresponds to the energy gap between the ground state and first excited state, because there are other relevant peaks.
However, it is worth mentioning that we can still obtain some information of the energy spectrum (or energy eigenvalues of the Hamiltonian) from the experimental data, even under the effect of the non-adiabatic
transitions between the ground state and other excited states. 
Again, this shows the robustness of our scheme against the non-adiabatic transitions compared with the previous schemes \cite{seki2020excited}.


\section{Conclusion}
In conclusion, we propose a scheme that
 allows the direct estimation of an energy gap 
of the target Hamiltonian by using quantum annealing (QA). While a ground state of a driving Hamiltonian 
is prepared as an initial state for the conventional QA, we prepare a superposition between a ground state 
and the first excited state of the driving Hamiltonian as the initial state. Also, the key idea in our scheme 
is to use a Ramsey type measurement after the quantum annealing process where information of the energy gap 
is encoded as a relative phase between the superposition. The readout of the relative phase by sweeping the 
Ramsey measurement time duration provides a direct estimation of the energy gap of the target Hamiltonian. 
We show that, unlike the previous scheme, our scheme is robust against non-adiabatic transitions.
Our 
scheme paves an alternative way to estimate the energy gap of the target Hamiltonian for applications of quantum chemistry.

While this manuscript was being written, 
an independent article also proposes to use a Ramsey measurement to estimate an energy 
gap by using a quanutm device \cite{2020quantumenergygap}.

 This paper is partly based on results obtained from a 
project commissioned by the New Energy and Industrial Technology Development Organization (NEDO), Japan. 
This work was also supported 
by Leading Initiative for Excellent Young Researchers MEXT Japan, JST presto (Grant No. JPMJPR1919) Japan
, KAKENHI Scientific Research C (Grant No. 18K03465), and JST-PRESTO (JPMJPR1914).


\begin{thebibliography}{34}%
\makeatletter
\providecommand \@ifxundefined [1]{%
 \@ifx{#1\undefined}
}%
\providecommand \@ifnum [1]{%
 \ifnum #1\expandafter \@firstoftwo
 \else \expandafter \@secondoftwo
 \fi
}%
\providecommand \@ifx [1]{%
 \ifx #1\expandafter \@firstoftwo
 \else \expandafter \@secondoftwo
 \fi
}%
\providecommand \natexlab [1]{#1}%
\providecommand \enquote  [1]{``#1''}%
\providecommand \bibnamefont  [1]{#1}%
\providecommand \bibfnamefont [1]{#1}%
\providecommand \citenamefont [1]{#1}%
\providecommand \href@noop [0]{\@secondoftwo}%
\providecommand \href [0]{\begingroup \@sanitize@url \@href}%
\providecommand \@href[1]{\@@startlink{#1}\@@href}%
\providecommand \@@href[1]{\endgroup#1\@@endlink}%
\providecommand \@sanitize@url [0]{\catcode `\\12\catcode `\$12\catcode
  `\&12\catcode `\#12\catcode `\^12\catcode `\_12\catcode `\%12\relax}%
\providecommand \@@startlink[1]{}%
\providecommand \@@endlink[0]{}%
\providecommand \url  [0]{\begingroup\@sanitize@url \@url }%
\providecommand \@url [1]{\endgroup\@href {#1}{\urlprefix }}%
\providecommand \urlprefix  [0]{URL }%
\providecommand \Eprint [0]{\href }%
\providecommand \doibase [0]{http://dx.doi.org/}%
\providecommand \selectlanguage [0]{\@gobble}%
\providecommand \bibinfo  [0]{\@secondoftwo}%
\providecommand \bibfield  [0]{\@secondoftwo}%
\providecommand \translation [1]{[#1]}%
\providecommand \BibitemOpen [0]{}%
\providecommand \bibitemStop [0]{}%
\providecommand \bibitemNoStop [0]{.\EOS\space}%
\providecommand \EOS [0]{\spacefactor3000\relax}%
\providecommand \BibitemShut  [1]{\csname bibitem#1\endcsname}%
\let\auto@bib@innerbib\@empty
\bibitem [{\citenamefont {Kadowaki}\ and\ \citenamefont
  {Nishimori}(1998)}]{kadowaki1998quantum}%
  \BibitemOpen
  \bibfield  {author} {\bibinfo {author} {\bibfnamefont {T.}~\bibnamefont
  {Kadowaki}}\ and\ \bibinfo {author} {\bibfnamefont {H.}~\bibnamefont
  {Nishimori}},\ }\href@noop {} {\bibfield  {journal} {\bibinfo  {journal}
  {Physical Review E}\ }\textbf {\bibinfo {volume} {58}},\ \bibinfo {pages}
  {5355} (\bibinfo {year} {1998})}\BibitemShut {NoStop}%
\bibitem [{\citenamefont {Farhi}\ \emph {et~al.}(2000)\citenamefont {Farhi},
  \citenamefont {Goldstone}, \citenamefont {Gutmann},\ and\ \citenamefont
  {Sipser}}]{farhi2000quantum}%
  \BibitemOpen
  \bibfield  {author} {\bibinfo {author} {\bibfnamefont {E.}~\bibnamefont
  {Farhi}}, \bibinfo {author} {\bibfnamefont {J.}~\bibnamefont {Goldstone}},
  \bibinfo {author} {\bibfnamefont {S.}~\bibnamefont {Gutmann}}, \ and\
  \bibinfo {author} {\bibfnamefont {M.}~\bibnamefont {Sipser}},\ }\href@noop {}
  {\bibfield  {journal} {\bibinfo  {journal} {arXiv preprint quant-ph/0001106}\
  } (\bibinfo {year} {2000})}\BibitemShut {NoStop}%
\bibitem [{\citenamefont {Farhi}\ \emph {et~al.}(2001)\citenamefont {Farhi},
  \citenamefont {Goldstone}, \citenamefont {Gutmann}, \citenamefont {Lapan},
  \citenamefont {Lundgren},\ and\ \citenamefont {Preda}}]{farhi2001quantum}%
  \BibitemOpen
  \bibfield  {author} {\bibinfo {author} {\bibfnamefont {E.}~\bibnamefont
  {Farhi}}, \bibinfo {author} {\bibfnamefont {J.}~\bibnamefont {Goldstone}},
  \bibinfo {author} {\bibfnamefont {S.}~\bibnamefont {Gutmann}}, \bibinfo
  {author} {\bibfnamefont {J.}~\bibnamefont {Lapan}}, \bibinfo {author}
  {\bibfnamefont {A.}~\bibnamefont {Lundgren}}, \ and\ \bibinfo {author}
  {\bibfnamefont {D.}~\bibnamefont {Preda}},\ }\href@noop {} {\bibfield
  {journal} {\bibinfo  {journal} {Science}\ }\textbf {\bibinfo {volume}
  {292}},\ \bibinfo {pages} {472} (\bibinfo {year} {2001})}\BibitemShut
  {NoStop}%
\bibitem [{\citenamefont {Johnson}\ \emph {et~al.}(2011)\citenamefont
  {Johnson}, \citenamefont {Amin}, \citenamefont {Gildert}, \citenamefont
  {Lanting}, \citenamefont {Hamze}, \citenamefont {Dickson}, \citenamefont
  {Harris}, \citenamefont {Berkley}, \citenamefont {Johansson}, \citenamefont
  {Bunyk} \emph {et~al.}}]{johnson2011quantum}%
  \BibitemOpen
  \bibfield  {author} {\bibinfo {author} {\bibfnamefont {M.~W.}\ \bibnamefont
  {Johnson}}, \bibinfo {author} {\bibfnamefont {M.~H.}\ \bibnamefont {Amin}},
  \bibinfo {author} {\bibfnamefont {S.}~\bibnamefont {Gildert}}, \bibinfo
  {author} {\bibfnamefont {T.}~\bibnamefont {Lanting}}, \bibinfo {author}
  {\bibfnamefont {F.}~\bibnamefont {Hamze}}, \bibinfo {author} {\bibfnamefont
  {N.}~\bibnamefont {Dickson}}, \bibinfo {author} {\bibfnamefont
  {R.}~\bibnamefont {Harris}}, \bibinfo {author} {\bibfnamefont {A.~J.}\
  \bibnamefont {Berkley}}, \bibinfo {author} {\bibfnamefont {J.}~\bibnamefont
  {Johansson}}, \bibinfo {author} {\bibfnamefont {P.}~\bibnamefont {Bunyk}},
  \emph {et~al.},\ }\href@noop {} {\bibfield  {journal} {\bibinfo  {journal}
  {Nature}\ }\textbf {\bibinfo {volume} {473}},\ \bibinfo {pages} {194}
  (\bibinfo {year} {2011})}\BibitemShut {NoStop}%
\bibitem [{\citenamefont {Orlando}\ \emph {et~al.}(1999)\citenamefont
  {Orlando}, \citenamefont {Mooij}, \citenamefont {Tian}, \citenamefont {Van
  Der~Wal}, \citenamefont {Levitov}, \citenamefont {Lloyd},\ and\ \citenamefont
  {Mazo}}]{orlando1999superconducting}%
  \BibitemOpen
  \bibfield  {author} {\bibinfo {author} {\bibfnamefont {T.}~\bibnamefont
  {Orlando}}, \bibinfo {author} {\bibfnamefont {J.}~\bibnamefont {Mooij}},
  \bibinfo {author} {\bibfnamefont {L.}~\bibnamefont {Tian}}, \bibinfo {author}
  {\bibfnamefont {C.~H.}\ \bibnamefont {Van Der~Wal}}, \bibinfo {author}
  {\bibfnamefont {L.}~\bibnamefont {Levitov}}, \bibinfo {author} {\bibfnamefont
  {S.}~\bibnamefont {Lloyd}}, \ and\ \bibinfo {author} {\bibfnamefont
  {J.}~\bibnamefont {Mazo}},\ }\href@noop {} {\bibfield  {journal} {\bibinfo
  {journal} {Physical Review B}\ }\textbf {\bibinfo {volume} {60}},\ \bibinfo
  {pages} {15398} (\bibinfo {year} {1999})}\BibitemShut {NoStop}%
\bibitem [{\citenamefont {Mooij}\ \emph {et~al.}(1999)\citenamefont {Mooij},
  \citenamefont {Orlando}, \citenamefont {Levitov}, \citenamefont {Tian},
  \citenamefont {Van~der Wal},\ and\ \citenamefont
  {Lloyd}}]{mooij1999josephson}%
  \BibitemOpen
  \bibfield  {author} {\bibinfo {author} {\bibfnamefont {J.}~\bibnamefont
  {Mooij}}, \bibinfo {author} {\bibfnamefont {T.}~\bibnamefont {Orlando}},
  \bibinfo {author} {\bibfnamefont {L.}~\bibnamefont {Levitov}}, \bibinfo
  {author} {\bibfnamefont {L.}~\bibnamefont {Tian}}, \bibinfo {author}
  {\bibfnamefont {C.~H.}\ \bibnamefont {Van~der Wal}}, \ and\ \bibinfo {author}
  {\bibfnamefont {S.}~\bibnamefont {Lloyd}},\ }\href@noop {} {\bibfield
  {journal} {\bibinfo  {journal} {Science}\ }\textbf {\bibinfo {volume}
  {285}},\ \bibinfo {pages} {1036} (\bibinfo {year} {1999})}\BibitemShut
  {NoStop}%
\bibitem [{\citenamefont {Harris}\ \emph {et~al.}(2010)\citenamefont {Harris},
  \citenamefont {Johansson}, \citenamefont {Berkley}, \citenamefont {Johnson},
  \citenamefont {Lanting}, \citenamefont {Han}, \citenamefont {Bunyk},
  \citenamefont {Ladizinsky}, \citenamefont {Oh}, \citenamefont {Perminov}
  \emph {et~al.}}]{harris2010experimental}%
  \BibitemOpen
  \bibfield  {author} {\bibinfo {author} {\bibfnamefont {R.}~\bibnamefont
  {Harris}}, \bibinfo {author} {\bibfnamefont {J.}~\bibnamefont {Johansson}},
  \bibinfo {author} {\bibfnamefont {A.}~\bibnamefont {Berkley}}, \bibinfo
  {author} {\bibfnamefont {M.}~\bibnamefont {Johnson}}, \bibinfo {author}
  {\bibfnamefont {T.}~\bibnamefont {Lanting}}, \bibinfo {author} {\bibfnamefont
  {S.}~\bibnamefont {Han}}, \bibinfo {author} {\bibfnamefont {P.}~\bibnamefont
  {Bunyk}}, \bibinfo {author} {\bibfnamefont {E.}~\bibnamefont {Ladizinsky}},
  \bibinfo {author} {\bibfnamefont {T.}~\bibnamefont {Oh}}, \bibinfo {author}
  {\bibfnamefont {I.}~\bibnamefont {Perminov}},  \emph {et~al.},\ }\href@noop
  {} {\bibfield  {journal} {\bibinfo  {journal} {Physical Review B}\ }\textbf
  {\bibinfo {volume} {81}},\ \bibinfo {pages} {134510} (\bibinfo {year}
  {2010})}\BibitemShut {NoStop}%
\bibitem [{\citenamefont {Harris}\ \emph {et~al.}(2018)\citenamefont {Harris},
  \citenamefont {Sato}, \citenamefont {Berkley}, \citenamefont {Reis},
  \citenamefont {Altomare}, \citenamefont {Amin}, \citenamefont {Boothby},
  \citenamefont {Bunyk}, \citenamefont {Deng}, \citenamefont {Enderud} \emph
  {et~al.}}]{harris2018phase}%
  \BibitemOpen
  \bibfield  {author} {\bibinfo {author} {\bibfnamefont {R.}~\bibnamefont
  {Harris}}, \bibinfo {author} {\bibfnamefont {Y.}~\bibnamefont {Sato}},
  \bibinfo {author} {\bibfnamefont {A.}~\bibnamefont {Berkley}}, \bibinfo
  {author} {\bibfnamefont {M.}~\bibnamefont {Reis}}, \bibinfo {author}
  {\bibfnamefont {F.}~\bibnamefont {Altomare}}, \bibinfo {author}
  {\bibfnamefont {M.}~\bibnamefont {Amin}}, \bibinfo {author} {\bibfnamefont
  {K.}~\bibnamefont {Boothby}}, \bibinfo {author} {\bibfnamefont
  {P.}~\bibnamefont {Bunyk}}, \bibinfo {author} {\bibfnamefont
  {C.}~\bibnamefont {Deng}}, \bibinfo {author} {\bibfnamefont {C.}~\bibnamefont
  {Enderud}},  \emph {et~al.},\ }\href@noop {} {\bibfield  {journal} {\bibinfo
  {journal} {Science}\ }\textbf {\bibinfo {volume} {361}},\ \bibinfo {pages}
  {162} (\bibinfo {year} {2018})}\BibitemShut {NoStop}%
\bibitem [{\citenamefont {King}\ \emph {et~al.}(2018)\citenamefont {King},
  \citenamefont {Carrasquilla}, \citenamefont {Raymond}, \citenamefont
  {Ozfidan}, \citenamefont {Andriyash}, \citenamefont {Berkley}, \citenamefont
  {Reis}, \citenamefont {Lanting}, \citenamefont {Harris}, \citenamefont
  {Altomare} \emph {et~al.}}]{king2018observation}%
  \BibitemOpen
  \bibfield  {author} {\bibinfo {author} {\bibfnamefont {A.~D.}\ \bibnamefont
  {King}}, \bibinfo {author} {\bibfnamefont {J.}~\bibnamefont {Carrasquilla}},
  \bibinfo {author} {\bibfnamefont {J.}~\bibnamefont {Raymond}}, \bibinfo
  {author} {\bibfnamefont {I.}~\bibnamefont {Ozfidan}}, \bibinfo {author}
  {\bibfnamefont {E.}~\bibnamefont {Andriyash}}, \bibinfo {author}
  {\bibfnamefont {A.}~\bibnamefont {Berkley}}, \bibinfo {author} {\bibfnamefont
  {M.}~\bibnamefont {Reis}}, \bibinfo {author} {\bibfnamefont {T.}~\bibnamefont
  {Lanting}}, \bibinfo {author} {\bibfnamefont {R.}~\bibnamefont {Harris}},
  \bibinfo {author} {\bibfnamefont {F.}~\bibnamefont {Altomare}},  \emph
  {et~al.},\ }\href@noop {} {\bibfield  {journal} {\bibinfo  {journal}
  {Nature}\ }\textbf {\bibinfo {volume} {560}},\ \bibinfo {pages} {456}
  (\bibinfo {year} {2018})}\BibitemShut {NoStop}%
\bibitem [{\citenamefont {Mott}\ \emph {et~al.}(2017)\citenamefont {Mott},
  \citenamefont {Job}, \citenamefont {Vlimant}, \citenamefont {Lidar},\ and\
  \citenamefont {Spiropulu}}]{mott2017solving}%
  \BibitemOpen
  \bibfield  {author} {\bibinfo {author} {\bibfnamefont {A.}~\bibnamefont
  {Mott}}, \bibinfo {author} {\bibfnamefont {J.}~\bibnamefont {Job}}, \bibinfo
  {author} {\bibfnamefont {J.-R.}\ \bibnamefont {Vlimant}}, \bibinfo {author}
  {\bibfnamefont {D.}~\bibnamefont {Lidar}}, \ and\ \bibinfo {author}
  {\bibfnamefont {M.}~\bibnamefont {Spiropulu}},\ }\href@noop {} {\bibfield
  {journal} {\bibinfo  {journal} {Nature}\ }\textbf {\bibinfo {volume} {550}},\
  \bibinfo {pages} {375} (\bibinfo {year} {2017})}\BibitemShut {NoStop}%
\bibitem [{\citenamefont {Amin}\ \emph {et~al.}(2018)\citenamefont {Amin},
  \citenamefont {Andriyash}, \citenamefont {Rolfe}, \citenamefont
  {Kulchytskyy},\ and\ \citenamefont {Melko}}]{amin2018quantum}%
  \BibitemOpen
  \bibfield  {author} {\bibinfo {author} {\bibfnamefont {M.~H.}\ \bibnamefont
  {Amin}}, \bibinfo {author} {\bibfnamefont {E.}~\bibnamefont {Andriyash}},
  \bibinfo {author} {\bibfnamefont {J.}~\bibnamefont {Rolfe}}, \bibinfo
  {author} {\bibfnamefont {B.}~\bibnamefont {Kulchytskyy}}, \ and\ \bibinfo
  {author} {\bibfnamefont {R.}~\bibnamefont {Melko}},\ }\href@noop {}
  {\bibfield  {journal} {\bibinfo  {journal} {Physical Review X}\ }\textbf
  {\bibinfo {volume} {8}},\ \bibinfo {pages} {021050} (\bibinfo {year}
  {2018})}\BibitemShut {NoStop}%
\bibitem [{\citenamefont {Levine}\ \emph {et~al.}(2009)\citenamefont {Levine},
  \citenamefont {Busch},\ and\ \citenamefont {Shull}}]{levine2009quantum}%
  \BibitemOpen
  \bibfield  {author} {\bibinfo {author} {\bibfnamefont {I.~N.}\ \bibnamefont
  {Levine}}, \bibinfo {author} {\bibfnamefont {D.~H.}\ \bibnamefont {Busch}}, \
  and\ \bibinfo {author} {\bibfnamefont {H.}~\bibnamefont {Shull}},\
  }\href@noop {} {\emph {\bibinfo {title} {Quantum chemistry}}},\ Vol.~\bibinfo
  {volume} {6}\ (\bibinfo  {publisher} {Pearson Prentice Hall Upper Saddle
  River, NJ},\ \bibinfo {year} {2009})\BibitemShut {NoStop}%
\bibitem [{\citenamefont {Serrano-Andr{\'e}s}\ and\ \citenamefont
  {Merch{\'a}n}(2005)}]{serrano2005quantum}%
  \BibitemOpen
  \bibfield  {author} {\bibinfo {author} {\bibfnamefont {L.}~\bibnamefont
  {Serrano-Andr{\'e}s}}\ and\ \bibinfo {author} {\bibfnamefont
  {M.}~\bibnamefont {Merch{\'a}n}},\ }\href@noop {} {\bibfield  {journal}
  {\bibinfo  {journal} {Journal of Molecular Structure: THEOCHEM}\ }\textbf
  {\bibinfo {volume} {729}},\ \bibinfo {pages} {99} (\bibinfo {year}
  {2005})}\BibitemShut {NoStop}%
\bibitem [{\citenamefont {McArdle}\ \emph {et~al.}(2020)\citenamefont
  {McArdle}, \citenamefont {Endo}, \citenamefont {Aspuru-Guzik}, \citenamefont
  {Benjamin},\ and\ \citenamefont {Yuan}}]{mcardle2020quantum}%
  \BibitemOpen
  \bibfield  {author} {\bibinfo {author} {\bibfnamefont {S.}~\bibnamefont
  {McArdle}}, \bibinfo {author} {\bibfnamefont {S.}~\bibnamefont {Endo}},
  \bibinfo {author} {\bibfnamefont {A.}~\bibnamefont {Aspuru-Guzik}}, \bibinfo
  {author} {\bibfnamefont {S.~C.}\ \bibnamefont {Benjamin}}, \ and\ \bibinfo
  {author} {\bibfnamefont {X.}~\bibnamefont {Yuan}},\ }\href@noop {} {\bibfield
   {journal} {\bibinfo  {journal} {Reviews of Modern Physics}\ }\textbf
  {\bibinfo {volume} {92}},\ \bibinfo {pages} {015003} (\bibinfo {year}
  {2020})}\BibitemShut {NoStop}%
\bibitem [{\citenamefont {Perdomo-Ortiz}\ \emph {et~al.}(2012)\citenamefont
  {Perdomo-Ortiz}, \citenamefont {Dickson}, \citenamefont {Drew-Brook},
  \citenamefont {Rose},\ and\ \citenamefont
  {Aspuru-Guzik}}]{perdomo2012finding}%
  \BibitemOpen
  \bibfield  {author} {\bibinfo {author} {\bibfnamefont {A.}~\bibnamefont
  {Perdomo-Ortiz}}, \bibinfo {author} {\bibfnamefont {N.}~\bibnamefont
  {Dickson}}, \bibinfo {author} {\bibfnamefont {M.}~\bibnamefont {Drew-Brook}},
  \bibinfo {author} {\bibfnamefont {G.}~\bibnamefont {Rose}}, \ and\ \bibinfo
  {author} {\bibfnamefont {A.}~\bibnamefont {Aspuru-Guzik}},\ }\href@noop {}
  {\bibfield  {journal} {\bibinfo  {journal} {Scientific Reports}\ }\textbf
  {\bibinfo {volume} {2}},\ \bibinfo {pages} {571} (\bibinfo {year}
  {2012})}\BibitemShut {NoStop}%
\bibitem [{\citenamefont {Aspuru-Guzik}\ \emph {et~al.}(2005)\citenamefont
  {Aspuru-Guzik}, \citenamefont {Dutoi}, \citenamefont {Love},\ and\
  \citenamefont {Head-Gordon}}]{aspuru2005simulated}%
  \BibitemOpen
  \bibfield  {author} {\bibinfo {author} {\bibfnamefont {A.}~\bibnamefont
  {Aspuru-Guzik}}, \bibinfo {author} {\bibfnamefont {A.~D.}\ \bibnamefont
  {Dutoi}}, \bibinfo {author} {\bibfnamefont {P.~J.}\ \bibnamefont {Love}}, \
  and\ \bibinfo {author} {\bibfnamefont {M.}~\bibnamefont {Head-Gordon}},\
  }\href@noop {} {\bibfield  {journal} {\bibinfo  {journal} {Science}\ }\textbf
  {\bibinfo {volume} {309}},\ \bibinfo {pages} {1704} (\bibinfo {year}
  {2005})}\BibitemShut {NoStop}%
\bibitem [{\citenamefont {Lanyon}\ \emph {et~al.}(2010)\citenamefont {Lanyon},
  \citenamefont {Whitfield}, \citenamefont {Gillett}, \citenamefont {Goggin},
  \citenamefont {Almeida}, \citenamefont {Kassal}, \citenamefont {Biamonte},
  \citenamefont {Mohseni}, \citenamefont {Powell}, \citenamefont {Barbieri}
  \emph {et~al.}}]{lanyon2010towards}%
  \BibitemOpen
  \bibfield  {author} {\bibinfo {author} {\bibfnamefont {B.~P.}\ \bibnamefont
  {Lanyon}}, \bibinfo {author} {\bibfnamefont {J.~D.}\ \bibnamefont
  {Whitfield}}, \bibinfo {author} {\bibfnamefont {G.~G.}\ \bibnamefont
  {Gillett}}, \bibinfo {author} {\bibfnamefont {M.~E.}\ \bibnamefont {Goggin}},
  \bibinfo {author} {\bibfnamefont {M.~P.}\ \bibnamefont {Almeida}}, \bibinfo
  {author} {\bibfnamefont {I.}~\bibnamefont {Kassal}}, \bibinfo {author}
  {\bibfnamefont {J.~D.}\ \bibnamefont {Biamonte}}, \bibinfo {author}
  {\bibfnamefont {M.}~\bibnamefont {Mohseni}}, \bibinfo {author} {\bibfnamefont
  {B.~J.}\ \bibnamefont {Powell}}, \bibinfo {author} {\bibfnamefont
  {M.}~\bibnamefont {Barbieri}},  \emph {et~al.},\ }\href@noop {} {\bibfield
  {journal} {\bibinfo  {journal} {Nature chemistry}\ }\textbf {\bibinfo
  {volume} {2}},\ \bibinfo {pages} {106} (\bibinfo {year} {2010})}\BibitemShut
  {NoStop}%
\bibitem [{\citenamefont {Du}\ \emph {et~al.}(2010)\citenamefont {Du},
  \citenamefont {Xu}, \citenamefont {Peng}, \citenamefont {Wang}, \citenamefont
  {Wu},\ and\ \citenamefont {Lu}}]{du2010nmr}%
  \BibitemOpen
  \bibfield  {author} {\bibinfo {author} {\bibfnamefont {J.}~\bibnamefont
  {Du}}, \bibinfo {author} {\bibfnamefont {N.}~\bibnamefont {Xu}}, \bibinfo
  {author} {\bibfnamefont {X.}~\bibnamefont {Peng}}, \bibinfo {author}
  {\bibfnamefont {P.}~\bibnamefont {Wang}}, \bibinfo {author} {\bibfnamefont
  {S.}~\bibnamefont {Wu}}, \ and\ \bibinfo {author} {\bibfnamefont
  {D.}~\bibnamefont {Lu}},\ }\href@noop {} {\bibfield  {journal} {\bibinfo
  {journal} {Physical review letters}\ }\textbf {\bibinfo {volume} {104}},\
  \bibinfo {pages} {030502} (\bibinfo {year} {2010})}\BibitemShut {NoStop}%
\bibitem [{\citenamefont {Peruzzo}\ \emph {et~al.}(2014)\citenamefont
  {Peruzzo}, \citenamefont {McClean}, \citenamefont {Shadbolt}, \citenamefont
  {Yung}, \citenamefont {Zhou}, \citenamefont {Love}, \citenamefont
  {Aspuru-Guzik},\ and\ \citenamefont {O’brien}}]{peruzzo2014variational}%
  \BibitemOpen
  \bibfield  {author} {\bibinfo {author} {\bibfnamefont {A.}~\bibnamefont
  {Peruzzo}}, \bibinfo {author} {\bibfnamefont {J.}~\bibnamefont {McClean}},
  \bibinfo {author} {\bibfnamefont {P.}~\bibnamefont {Shadbolt}}, \bibinfo
  {author} {\bibfnamefont {M.-H.}\ \bibnamefont {Yung}}, \bibinfo {author}
  {\bibfnamefont {X.-Q.}\ \bibnamefont {Zhou}}, \bibinfo {author}
  {\bibfnamefont {P.~J.}\ \bibnamefont {Love}}, \bibinfo {author}
  {\bibfnamefont {A.}~\bibnamefont {Aspuru-Guzik}}, \ and\ \bibinfo {author}
  {\bibfnamefont {J.~L.}\ \bibnamefont {O’brien}},\ }\href@noop {} {\bibfield
   {journal} {\bibinfo  {journal} {Nature communications}\ }\textbf {\bibinfo
  {volume} {5}},\ \bibinfo {pages} {4213} (\bibinfo {year} {2014})}\BibitemShut
  {NoStop}%
\bibitem [{\citenamefont {Mazzola}\ \emph {et~al.}(2017)\citenamefont
  {Mazzola}, \citenamefont {Smelyanskiy},\ and\ \citenamefont
  {Troyer}}]{mazzola2017quantum}%
  \BibitemOpen
  \bibfield  {author} {\bibinfo {author} {\bibfnamefont {G.}~\bibnamefont
  {Mazzola}}, \bibinfo {author} {\bibfnamefont {V.~N.}\ \bibnamefont
  {Smelyanskiy}}, \ and\ \bibinfo {author} {\bibfnamefont {M.}~\bibnamefont
  {Troyer}},\ }\href@noop {} {\bibfield  {journal} {\bibinfo  {journal}
  {Physical Review B}\ }\textbf {\bibinfo {volume} {96}},\ \bibinfo {pages}
  {134305} (\bibinfo {year} {2017})}\BibitemShut {NoStop}%
\bibitem [{\citenamefont {Streif}\ \emph {et~al.}(2019)\citenamefont {Streif},
  \citenamefont {Neukart},\ and\ \citenamefont {Leib}}]{streif2019solving}%
  \BibitemOpen
  \bibfield  {author} {\bibinfo {author} {\bibfnamefont {M.}~\bibnamefont
  {Streif}}, \bibinfo {author} {\bibfnamefont {F.}~\bibnamefont {Neukart}}, \
  and\ \bibinfo {author} {\bibfnamefont {M.}~\bibnamefont {Leib}},\ }in\
  \href@noop {} {\emph {\bibinfo {booktitle} {International Workshop on Quantum
  Technology and Optimization Problems}}}\ (\bibinfo {organization}
  {Springer},\ \bibinfo {year} {2019})\ pp.\ \bibinfo {pages}
  {111--122}\BibitemShut {NoStop}%
\bibitem [{\citenamefont {Babbush}\ \emph {et~al.}(2014)\citenamefont
  {Babbush}, \citenamefont {Love},\ and\ \citenamefont
  {Aspuru-Guzik}}]{babbush2014adiabatic}%
  \BibitemOpen
  \bibfield  {author} {\bibinfo {author} {\bibfnamefont {R.}~\bibnamefont
  {Babbush}}, \bibinfo {author} {\bibfnamefont {P.~J.}\ \bibnamefont {Love}}, \
  and\ \bibinfo {author} {\bibfnamefont {A.}~\bibnamefont {Aspuru-Guzik}},\
  }\href@noop {} {\bibfield  {journal} {\bibinfo  {journal} {Scientific
  Reports}\ }\textbf {\bibinfo {volume} {4}},\ \bibinfo {pages} {6603}
  (\bibinfo {year} {2014})}\BibitemShut {NoStop}%
\bibitem [{\citenamefont {Jordan}\ and\ \citenamefont
  {Wigner}(1928)}]{jordanwigner}%
  \BibitemOpen
  \bibfield  {author} {\bibinfo {author} {\bibfnamefont {P.}~\bibnamefont
  {Jordan}}\ and\ \bibinfo {author} {\bibfnamefont {E.}~\bibnamefont
  {Wigner}},\ }\href@noop {} {\bibfield  {journal} {\bibinfo  {journal} {Z.
  Physik}\ }\textbf {\bibinfo {volume} {47}},\ \bibinfo {pages} {631} (\bibinfo
  {year} {1928})}\BibitemShut {NoStop}%
\bibitem [{\citenamefont {Bravyi}\ and\ \citenamefont
  {Kitaev}(2002)}]{bravyi2002fermionic}%
  \BibitemOpen
  \bibfield  {author} {\bibinfo {author} {\bibfnamefont {S.~B.}\ \bibnamefont
  {Bravyi}}\ and\ \bibinfo {author} {\bibfnamefont {A.~Y.}\ \bibnamefont
  {Kitaev}},\ }\href@noop {} {\bibfield  {journal} {\bibinfo  {journal} {Annals
  of Physics}\ }\textbf {\bibinfo {volume} {298}},\ \bibinfo {pages} {210}
  (\bibinfo {year} {2002})}\BibitemShut {NoStop}%
\bibitem [{\citenamefont {Seeley}\ \emph {et~al.}(2012)\citenamefont {Seeley},
  \citenamefont {Richard},\ and\ \citenamefont {Love}}]{seeley2012bravyi}%
  \BibitemOpen
  \bibfield  {author} {\bibinfo {author} {\bibfnamefont {J.~T.}\ \bibnamefont
  {Seeley}}, \bibinfo {author} {\bibfnamefont {M.~J.}\ \bibnamefont {Richard}},
  \ and\ \bibinfo {author} {\bibfnamefont {P.~J.}\ \bibnamefont {Love}},\
  }\href@noop {} {\bibfield  {journal} {\bibinfo  {journal} {The Journal of
  Chemical physics}\ }\textbf {\bibinfo {volume} {137}},\ \bibinfo {pages}
  {224109} (\bibinfo {year} {2012})}\BibitemShut {NoStop}%
\bibitem [{\citenamefont {Tranter}\ \emph {et~al.}(2015)\citenamefont
  {Tranter}, \citenamefont {Sofia}, \citenamefont {Seeley}, \citenamefont
  {Kaicher}, \citenamefont {McClean}, \citenamefont {Babbush}, \citenamefont
  {Coveney}, \citenamefont {Mintert}, \citenamefont {Wilhelm},\ and\
  \citenamefont {Love}}]{tranter2015b}%
  \BibitemOpen
  \bibfield  {author} {\bibinfo {author} {\bibfnamefont {A.}~\bibnamefont
  {Tranter}}, \bibinfo {author} {\bibfnamefont {S.}~\bibnamefont {Sofia}},
  \bibinfo {author} {\bibfnamefont {J.}~\bibnamefont {Seeley}}, \bibinfo
  {author} {\bibfnamefont {M.}~\bibnamefont {Kaicher}}, \bibinfo {author}
  {\bibfnamefont {J.}~\bibnamefont {McClean}}, \bibinfo {author} {\bibfnamefont
  {R.}~\bibnamefont {Babbush}}, \bibinfo {author} {\bibfnamefont {P.~V.}\
  \bibnamefont {Coveney}}, \bibinfo {author} {\bibfnamefont {F.}~\bibnamefont
  {Mintert}}, \bibinfo {author} {\bibfnamefont {F.}~\bibnamefont {Wilhelm}}, \
  and\ \bibinfo {author} {\bibfnamefont {P.~J.}\ \bibnamefont {Love}},\
  }\href@noop {} {\bibfield  {journal} {\bibinfo  {journal} {International
  Journal of Quantum Chemistry}\ }\textbf {\bibinfo {volume} {115}},\ \bibinfo
  {pages} {1431} (\bibinfo {year} {2015})}\BibitemShut {NoStop}%
\bibitem [{\citenamefont {Chen}\ \emph {et~al.}(2019)\citenamefont {Chen},
  \citenamefont {Gong}, \citenamefont {Xu}, \citenamefont {Yuan}, \citenamefont
  {Wang}, \citenamefont {Wang}, \citenamefont {Ying}, \citenamefont {Lin},
  \citenamefont {Xu}, \citenamefont {Wu} \emph
  {et~al.}}]{chen2019demonstration}%
  \BibitemOpen
  \bibfield  {author} {\bibinfo {author} {\bibfnamefont {M.-C.}\ \bibnamefont
  {Chen}}, \bibinfo {author} {\bibfnamefont {M.}~\bibnamefont {Gong}}, \bibinfo
  {author} {\bibfnamefont {X.-S.}\ \bibnamefont {Xu}}, \bibinfo {author}
  {\bibfnamefont {X.}~\bibnamefont {Yuan}}, \bibinfo {author} {\bibfnamefont
  {J.-W.}\ \bibnamefont {Wang}}, \bibinfo {author} {\bibfnamefont
  {C.}~\bibnamefont {Wang}}, \bibinfo {author} {\bibfnamefont {C.}~\bibnamefont
  {Ying}}, \bibinfo {author} {\bibfnamefont {J.}~\bibnamefont {Lin}}, \bibinfo
  {author} {\bibfnamefont {Y.}~\bibnamefont {Xu}}, \bibinfo {author}
  {\bibfnamefont {Y.}~\bibnamefont {Wu}},  \emph {et~al.},\ }\href@noop {}
  {\bibfield  {journal} {\bibinfo  {journal} {arXiv preprint arXiv:1905.03150}\
  } (\bibinfo {year} {2019})}\BibitemShut {NoStop}%
\bibitem [{\citenamefont {Seki}\ \emph {et~al.}(2020)\citenamefont {Seki},
  \citenamefont {Matsuzaki},\ and\ \citenamefont {Kawabata}}]{seki2020excited}%
  \BibitemOpen
  \bibfield  {author} {\bibinfo {author} {\bibfnamefont {Y.}~\bibnamefont
  {Seki}}, \bibinfo {author} {\bibfnamefont {Y.}~\bibnamefont {Matsuzaki}}, \
  and\ \bibinfo {author} {\bibfnamefont {S.}~\bibnamefont {Kawabata}},\
  }\href@noop {} {\bibfield  {journal} {\bibinfo  {journal} {arXiv preprint
  arXiv:2002.12621}\ } (\bibinfo {year} {2020})}\BibitemShut {NoStop}%
\bibitem [{\citenamefont {Ramsey}(1950)}]{ramsey1950molecular}%
  \BibitemOpen
  \bibfield  {author} {\bibinfo {author} {\bibfnamefont {N.~F.}\ \bibnamefont
  {Ramsey}},\ }\href@noop {} {\bibfield  {journal} {\bibinfo  {journal}
  {Physical Review}\ }\textbf {\bibinfo {volume} {78}},\ \bibinfo {pages} {695}
  (\bibinfo {year} {1950})}\BibitemShut {NoStop}%
\bibitem [{\citenamefont {Perdomo-Ortiz}\ \emph {et~al.}(2011)\citenamefont
  {Perdomo-Ortiz}, \citenamefont {Venegas-Andraca},\ and\ \citenamefont
  {Aspuru-Guzik}}]{perdomo2011study}%
  \BibitemOpen
  \bibfield  {author} {\bibinfo {author} {\bibfnamefont {A.}~\bibnamefont
  {Perdomo-Ortiz}}, \bibinfo {author} {\bibfnamefont {S.~E.}\ \bibnamefont
  {Venegas-Andraca}}, \ and\ \bibinfo {author} {\bibfnamefont {A.}~\bibnamefont
  {Aspuru-Guzik}},\ }\href@noop {} {\bibfield  {journal} {\bibinfo  {journal}
  {Quantum Information Processing}\ }\textbf {\bibinfo {volume} {10}},\
  \bibinfo {pages} {33} (\bibinfo {year} {2011})}\BibitemShut {NoStop}%
\bibitem [{\citenamefont {Ohkuwa}\ \emph {et~al.}(2018)\citenamefont {Ohkuwa},
  \citenamefont {Nishimori},\ and\ \citenamefont {Lidar}}]{ohkuwa2018reverse}%
  \BibitemOpen
  \bibfield  {author} {\bibinfo {author} {\bibfnamefont {M.}~\bibnamefont
  {Ohkuwa}}, \bibinfo {author} {\bibfnamefont {H.}~\bibnamefont {Nishimori}}, \
  and\ \bibinfo {author} {\bibfnamefont {D.~A.}\ \bibnamefont {Lidar}},\
  }\href@noop {} {\bibfield  {journal} {\bibinfo  {journal} {Physical Review
  A}\ }\textbf {\bibinfo {volume} {98}},\ \bibinfo {pages} {022314} (\bibinfo
  {year} {2018})}\BibitemShut {NoStop}%
\bibitem [{\citenamefont {Yamashiro}\ \emph {et~al.}(2019)\citenamefont
  {Yamashiro}, \citenamefont {Ohkuwa}, \citenamefont {Nishimori},\ and\
  \citenamefont {Lidar}}]{yamashiro2019dynamics}%
  \BibitemOpen
  \bibfield  {author} {\bibinfo {author} {\bibfnamefont {Y.}~\bibnamefont
  {Yamashiro}}, \bibinfo {author} {\bibfnamefont {M.}~\bibnamefont {Ohkuwa}},
  \bibinfo {author} {\bibfnamefont {H.}~\bibnamefont {Nishimori}}, \ and\
  \bibinfo {author} {\bibfnamefont {D.~A.}\ \bibnamefont {Lidar}},\ }\href@noop
  {} {\bibfield  {journal} {\bibinfo  {journal} {Physical Review A}\ }\textbf
  {\bibinfo {volume} {100}},\ \bibinfo {pages} {052321} (\bibinfo {year}
  {2019})}\BibitemShut {NoStop}%
\bibitem [{\citenamefont {Arai}\ \emph {et~al.}(2020)\citenamefont {Arai},
  \citenamefont {Ohzeki},\ and\ \citenamefont {Tanaka}}]{arai2020mean}%
  \BibitemOpen
  \bibfield  {author} {\bibinfo {author} {\bibfnamefont {S.}~\bibnamefont
  {Arai}}, \bibinfo {author} {\bibfnamefont {M.}~\bibnamefont {Ohzeki}}, \ and\
  \bibinfo {author} {\bibfnamefont {K.}~\bibnamefont {Tanaka}},\ }\href@noop {}
  {\bibfield  {journal} {\bibinfo  {journal} {arXiv preprint arXiv:2004.11066}\
  } (\bibinfo {year} {2020})}\BibitemShut {NoStop}%
\bibitem [{\citenamefont {Russo}\ \emph {et~al.}(2020)\citenamefont {Russo},
  \citenamefont {Rudinger}, \citenamefont {Morrison},\ and\ \citenamefont
  {Baczewski}}]{2020quantumenergygap}%
  \BibitemOpen
  \bibfield  {author} {\bibinfo {author} {\bibfnamefont {A.~E.}\ \bibnamefont
  {Russo}}, \bibinfo {author} {\bibfnamefont {K.~M.}\ \bibnamefont {Rudinger}},
  \bibinfo {author} {\bibfnamefont {B.~C.~A.}\ \bibnamefont {Morrison}}, \ and\
  \bibinfo {author} {\bibfnamefont {A.~D.}\ \bibnamefont {Baczewski}},\
  }\href@noop {} {\bibfield  {journal} {\bibinfo  {journal} {arXiv:2007.08697}\
  } (\bibinfo {year} {2020})}\BibitemShut {NoStop}%
\end{thebibliography}
%

\end{document}